\documentclass[conference]{IEEEtran}
\IEEEoverridecommandlockouts
\usepackage{cite}
\usepackage{amsmath,amssymb,amsfonts}

\usepackage{caption}
\usepackage{colortbl}
\usepackage{booktabs}
\usepackage{algorithmic}
\usepackage{hyperref}
\usepackage{graphicx}
\usepackage{textcomp}
\usepackage{xcolor}
\usepackage{xurl}
\usepackage[ruled,linesnumbered]{algorithm2e}
\usepackage{multirow}
\usepackage{tcolorbox}

\usepackage{enumitem}
\def\BibTeX{{\rm B\kern-.05em{\sc i\kern-.025em b}\kern-.08em
    T\kern-.1667em\lower.7ex\hbox{E}\kern-.125emX}}

\makeatletter
\newcommand{\linebreakand}{%
  \end{@IEEEauthorhalign}
  \hfill\mbox{}\par
  \mbox{}\hfill\begin{@IEEEauthorhalign}
}
\makeatother

\begin{document}
\bibliographystyle{plain}

\title{
Learning Graph-based Patch Representations for Identifying and Assessing Silent Vulnerability Fixes
\\
}

% \author{\IEEEauthorblockN{1\textsuperscript{st} Mei Han}
% \IEEEauthorblockA{\textit{College of Software Engineering} \\
% \textit{Southeast University}\\
% Nanjing, China \\
% hanmei@seu.edu.cn}
% \and
% \IEEEauthorblockN{2\textsuperscript{nd} Lulu Wang$^{\ast}$}
% \thanks{*Corresponding author}
% \IEEEauthorblockA{\textit{School of Computer Science and} \\
% \textit{Engineering} \\
% \textit{Southeast University}\\
% Nanjing, China \\
% wanglulu@seu.edu.cn}
% \and
% \IEEEauthorblockN{3\textsuperscript{rd} Jianming Chang}
% \IEEEauthorblockA{\textit{School of Computer Science and} \\
% \textit{Engineering} \\
% \textit{Southeast University}\\
% Nanjing, China \\
% jianmingchang@seu.edu.cn}
% \linebreakand
% \IEEEauthorblockN{4\textsuperscript{th} Bixin Li}
% \IEEEauthorblockA{\textit{School of Computer Science and} \\
% \textit{Engineering} \\
% \textit{Southeast University}\\
% Nanjing, China \\
% bx.li@seu.edu.cn}
% \and
% \IEEEauthorblockN{5\textsuperscript{th} Chunguang Zhang}
% \IEEEauthorblockA{\textit{School of Computer Science and} \\
% \textit{Engineering} \\
% \textit{Southeast University}\\
% Nanjing, China \\
% 18905107369@163.com}
% }
\author{
    \IEEEauthorblockN{Mei Han, Lulu Wang$^\ast$, Jianming Chang, Bixin Li, Chunguang Zhang}
    \thanks{*Corresponding author}
    \IEEEauthorblockA{Southeast University, China}
    \IEEEauthorblockA{\{hanmei, wanglulu, jianmingchang, bx.li\}@seu.edu.cn, 18905107369@163.com}
}

\maketitle

\newcommand{\name}{GRAPE\xspace}
\begin{abstract}

Software projects are dependent on many third-party libraries, therefore high-risk vulnerabilities can propagate through the dependency chain to downstream projects. Owing to the subjective nature of patch management, software vendors commonly fix vulnerabilities silently. Silent vulnerability fixes cause downstream software to be unaware of urgent security issues in a timely manner, posing a security risk to the software. 
Presently, most of the existing works for vulnerability fix identification only consider the changed code as a sequential textual sequence, ignoring the structural information of the code.
 
In this paper, we propose \name, a GRAph-based Patch rEpresentation that aims to 1) provide a unified framework for getting vulnerability fix patches representation; and 2) enhance the understanding of the intent and potential impact of patches by extracting structural information of the code.
\name employs a novel joint graph structure (MCPG) to represent the syntactic and semantic information of silent fix patches and embeds both nodes and edges. Subsequently, a carefully designed graph convolutional neural network (NE-GCN) is utilized to fully learn structural features by leveraging the attributes of the nodes and edges.
Moreover, we construct a dataset containing 2251 silent fixes. For the experimental section, we evaluated patch representation on three tasks, including vulnerability fix identification, vulnerability types classification, and vulnerability severity classification. Experimental results indicate that, in comparison to baseline methods, \name can more effectively reduce false positives and omissions of vulnerability fixes identification and provide accurate vulnerability assessments.

\end{abstract}


\section{Introduction}

%第一段
%With the continuous development of the open source software (OSS) ecosystem, the development model relying on third-party repositories is becoming mainstream, gradually forming a software supply chain. Public research reports that 96\% of code bases contain third-party open source code\cite{r57}. However, this development model inevitably leads to the propagation of high-risk vulnerabilities to other projects through the dependency chain, thus expanding the impact of vulnerabilities.For example, Log4Shell \cite{r10} \cite{r8} is a 0-day vulnerability in the widely used Java logging repository Log4j. Log4Shell has affected more than 35,000 downstream Java package repositories, which has attracted a tremendous attention in the field of software system security. \cite{r9}.Therefore, it is crucial to sense and know about vulnerability fixes in upstream software repositories as early as possible.
Modern software projects commonly rely on a large amount of upstream code, with public studies reporting that 96\% of open source software (OSS) contain third-party libraries \cite{r57}. According to a recent study, vulnerabilities in upstream software are typically retained in downstream software for a long time. The delay in patch propagation leads to the propagation of high-risk vulnerabilities to other projects through the dependency chain, thus expanding the impact of vulnerabilities. For example, the remote command execution vulnerability (CVE-2021-22205 \cite{r71}) was disclosed in April 2021. However, seven months after its release, over 30,000 unpatched GitLab servers were hacked. Therefore, downstream software should be aware of vulnerability fixes in upstream software as early as possible.

%第二段
%Monitoring security notices of vulnerability databases is the most common method used by OSS maintainers to enhance supply chain security.
%The National Vulnerability Database (NVD) is the most popular, which built on and fully synchronized with the Common Vulnerabilities and Exposures (CVE) list, with additional details information including CWE types, severity scores, and more \cite{r14}.
%Unfortunately, this methodology faces two main problems. For one thing, there is commonly a time delay between fixing the vulnerability and reporting it, and studies have found that more than half of CVEs are fixed more than a week before they are publicly disclosed \cite{r13}. This means that a malicious party can seize this time gap to exploit the vulnerability to attack the downstream software of open source libraries. For another thing, the coverage of NVD is relatively low. According to a study, only 47\% of OSS vulnerabilities are disclosed through CVEs \cite{r12}, which suggests that most of the vulnerabilities are silently fixed without being publicly disclosed in any official reports. 
%So, it is necessary to propose a way to automatically identify vulnerability fixes.

Software maintainers tend to refer to security advisories, like Common Vulnerabilities and Exposures (CVE) \cite{r76}, to become aware of vulnerabilities and their fixes in upstream software.
When software fixes a vulnerability, it publicly discloses the vulnerability and assigns a CVE, and the vulnerability is recorded in the public vulnerability database, such as the National Vulnerability Database (NVD) \cite{r73}. In reality, due to the concerns about the quality reputation of software and easy software development management, some software vendors choose to release vulnerability fixes on the down-low \cite{r75}, \cite{r72}. These vulnerability fixes (i.e., push commits to the codebase without reporting to the NVD or explicitly indicating the vulnerability in log messages) are referred to as silent vulnerability fixes \cite{r34}. Silent vulnerability fixes pose a conundrum for downstream software maintainers, leaving them in the dark about the true security implications and the urgency to apply them. Consequently, it is necessary to propose an automatic vulnerability fix identification method for tracking upstream fix behavior.

%第三段
%Researchers are presently proposing many approaches based on deep learning to automatically identify vulnerability fixes. There are three main groups of these approaches, which are mainly divided into three categories: those based on recurrent neural networks (RNNs) \cite{r6} \cite{r23}, based on graph neural networks (GNNs) \cite{r4} \cite{r12}, and based on large Language models (LLMs) \cite{r20} \cite{r21} \cite{r24}. However, these methods have some limitations, such as RNNs ignores structural features of the code, and LLMs requires a large amount of data support. These factors lead to a high miss rate in existing research. Moveover, they often lack subsequent work to assess vulnerabilities.
%After filtering out the vulnerability fixes, understanding and analyzing vulnerability behavior can also be very time consuming and difficult. However, providing some information about the vulnerability can help OSS maintainers to better assess the vulnerability.
%CoLeFunDa \cite{r17} is the state-of-the-art (SOTA) early vulnerability fix perception approach, which can further explain the vulnerability after identifying the vulnerability fix. It is a function-level detection framework, and vulnerability fixes usually involve many functions in real-world applications, which prevents CoLeFunDa from adequately capturing the full view of the fix.
Researchers have proposed a number of learning-based approaches for vulnerability fix identification. Some utilize Machine Learning (ML) techniques that emphasize syntax features of the code \cite{r74}, \cite{r39}, while others utilize Recurrent Neural Networks (RNNs) that consider the code as a flat sequence of tokens \cite{r23}, \cite{r6}. However, ML-based methods only focus on the extraction of key features and lack the dependencies between code statements. Meanwhile, RNN-based methods treat changed code as sequential text sequences and lack structural features of the code. Inadequate code structure feature extraction leads to high miss rates of these methods, severely limiting their application.

Even if a vulnerability fix has been identified, understanding and analyzing vulnerability behavior can also be very time-consuming and difficult with manpower. It is generally recognized that providing additional vulnerability information to software maintainers is critical for them to prioritize fixes. CoLeFunDa \cite{r17} is the state-of-the-art early vulnerability fix perception approach, which can further explain the vulnerability after identifying the vulnerability fix. It is a function-level detection framework, and vulnerability fixes usually involve many functions in real-world applications, which prevents CoLeFunDa from adequately capturing the full view of the fix.

To deal with the above problems, our research focuses on extracting structural features of the code and further explaining the identified vulnerabilities. Typically, vulnerabilities are fixed in the form of security patches. To this end, we propose an innovative graph-based representation specialized for patches. This method is capable of merging the separately generated code property graphs (CPGs) of defect code and fixed code into a unified graph (MCPG). In this way, we preserve the important information in the graph while also reducing noise interference. We also design an advanced graph convolutional neural network (NE-GCN) that performs message-passing and aggregation operations on the feature vectors of nodes and edges to maximize the capture of relational information between nodes.

To evaluate the effectiveness of patch representations, we designed three experiments related to vulnerability fixes, namely silent vulnerability fix identification, vulnerability types classification, and vulnerability severity classification.
Vulnerability types and vulnerability severity are two common information used to help parse the behavior of the vulnerability. Common Weakness Enumeration (CWE) \cite{r15} is a standardized list used to identify and classify software vulnerabilities. By further dividing fix patches into fine-grained CWE types, it can help maintainers quickly determine the type of vulnerability, and thereby understand the behavior and characteristics of the vulnerability. Common Vulnerability Scoring System (CVSS) \cite{r16} is a widely used evaluation standard for characterizing the features and severity of software vulnerabilities. The vulnerability severity can help OSS maintainers properly determine and prioritize vulnerability risk. For example, the severity of Log4Shell \cite{r10}, \cite{r8} is critical, which falls into the most urgent category of vulnerabilities. OSS maintainers need to react to such vulnerabilities as soon as they receive the alert.

We evaluate GRAPE on the three tasks described above. We use a dataset consisting of 1068 vulnerability fix commits and 1183 non-vulnerability fix commits from Java OSS. We compare \name to six state-of-the-art deep learning methods, including three neural network models \cite{r1} \cite{r3} \cite{r2} widely used in natural language processing and three leading-edge methods \cite{r4} \cite{r5} \cite{r6} for silent vulnerability fixes. The experimental results show that our method outperforms all six baseline methods in terms of each metric. In the silent vulnerability fix identification task, \name outperforms the state-of-the-art baseline method by 7.1\% and 6.55\% in terms of accuracy and F1-score. In the vulnerability types multi-classification task, \name outperforms the state-of-the-art baseline method by 10.26\% and 12.57\% in terms of MCC and F1-score. In the vulnerability severity multi-classification task, \name outperformed the state-of-the-art baseline method by 14.19\% and 14.98\% in terms of MCC and F1-score.

In summary, the contributions of this paper can be summarized as follows:

\begin{itemize}
\item We propose MCPG, a graph representation applicable to patches, which combines syntactic and semantic information of silent fix patches, as well as the dependency and control relationships in the source code.
\item We design a graph convolutional neural network (NE-GCN) specialized for MCPG, which comprehensively captures the implicit features in silent fixes by passing and aggregating node features and edge features of the graph.
\item We evaluate \name using a constructed dataset on three tasks, including identifying vulnerability fixes, classifying vulnerability types, and classifying vulnerability severity. Experimental results show that \name achieves the best performance compared to the baselines.
\end{itemize}

\section{Background}

\subsection{Common Vulnerabilities and Exposure (CVE) and the National Vulnerability Database (NVD)}

CVE is a publicly disclosed database of known security vulnerabilities. Each entry in the CVE database is assigned a unique identifier, commonly referred to as a CVE-ID (e.g., (CVE-2021-22205 \cite{r71}). These identifiers facilitate accurate communication and discussion among security researchers, software vendors, and users about specific security vulnerabilities. The CVE database, maintained by the MITRE Corporation, is widely used in vulnerability management, vulnerability scanning tools, and security advisories.

NVD is a CVE vulnerability database maintained by the National Institute of Standards and Technology (NIST) in the United States.  Beyond just listing CVE entries, the NVD enriches the data with additional information such as CWE and CVSS scores, providing users with a more comprehensive understanding of each vulnerability's nature and severity. This additional context and the tools provided by the NVD make it a more robust resource for security professionals and organizations seeking to manage and mitigate potential risks in their systems.

\begin{figure*}[t]
  \begin{center}
    \includegraphics[width=0.92\textwidth]{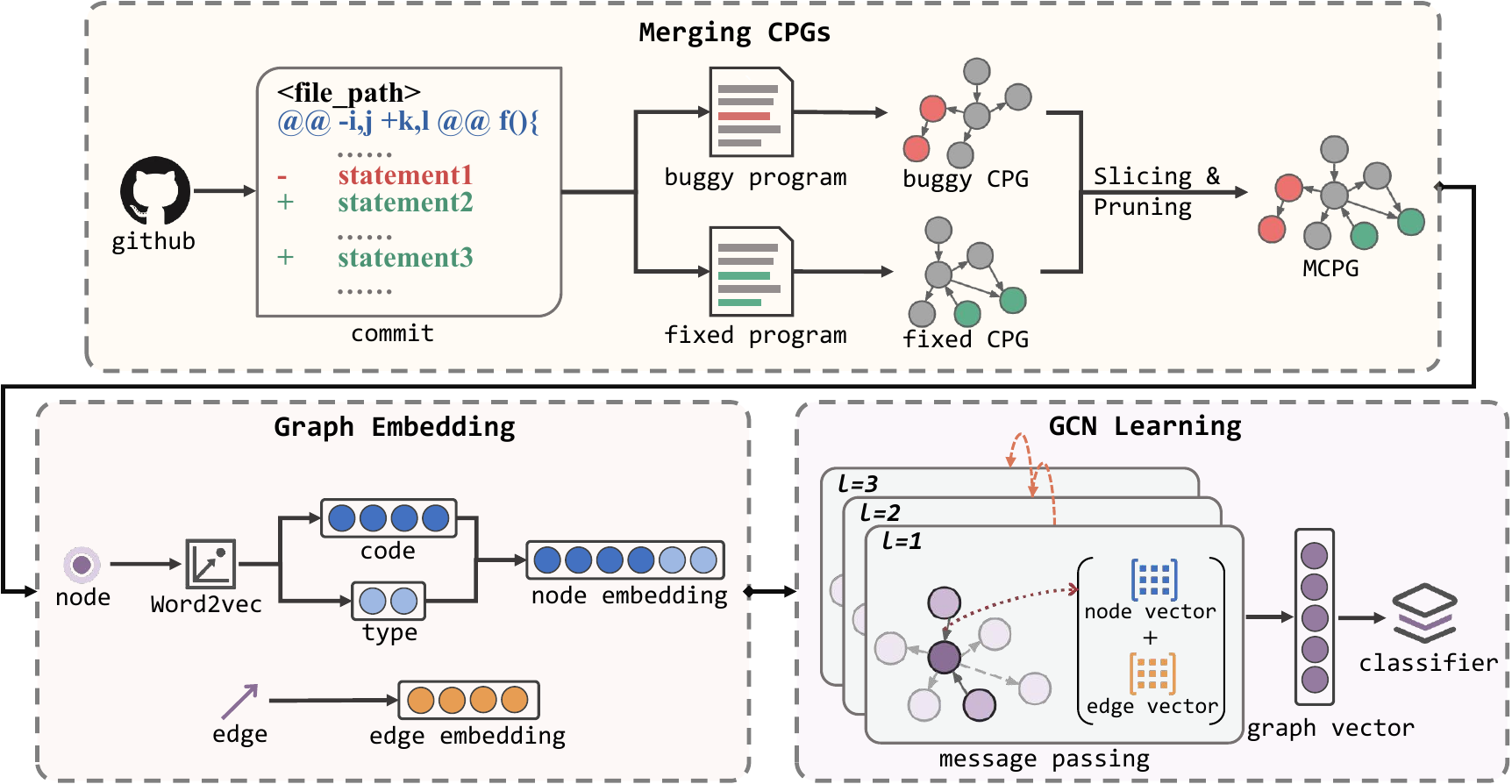}
    \caption{Overall Architecture of \name. }
    \label{f1}
  \end{center}
\end{figure*}

\subsection{Common Weakness Enumeration (CWE) and Common Vulnerability Scoring System (CVSS)}

CWE is an authoritative, community-driven list that describes security weaknesses in software and hardware. It provides a standardized system for classifying and describing vulnerabilities, enabling security professionals to communicate, remediate, and defend against these weaknesses more effectively. Each weakness in the CWE list is assigned a unique identifier. The goal of CWE is to provide a common language and framework for better understanding and addressing security issues.

CVSS is an open standard for evaluating and scoring the severity of known information security vulnerabilities. It provides a method for quantifying the potential impact of a vulnerability, including threats to confidentiality, integrity, and availability. CVSS scores are calculated based on several attributes and typically range from 0 to 10, with higher scores meaning more severe vulnerabilities. CVSS aids security teams in prioritizing and remediating the most critical vulnerabilities, thereby allocating security resources more effectively. We use CVSS version 3.0 in our study and classify the severity of vulnerability fixes into four classes based on this version.

\subsection{Code Property Graph (CPG)}

Code Property Graph (CPG) is an intermediate representation that transcends programming languages, integrating various abstract forms of source code into a unified and searchable graph database. The CPG amalgamates three fundamental compiler representations: Abstract Syntax Tree (AST), Control Flow Graph (CFG), and Program Dependence Graph (PDG).
AST, generated through the compiler's syntactic analysis, illustrates the hierarchical structure of the code. CFG presents all potential paths traversed during program execution in a graphical manner. PDG encompasses both control and data dependencies within a program, represented through the Control Dependence Graph (CDG) and the Data Dependence Graph (DDG).

The comprehensiveness of the CPG lies in its integration of multi-dimensional information such as control flow, control dependencies, data dependencies, and program syntax, providing an in-depth perspective for static code analysis. CPG-based methods have been developed for vulnerability-related tasks. We merge the CPGs of buggy programs and fixed programs into a unified graph structure that can represent the structure of patches more comprehensively.

\section{Approach}
In this section, we first introduce an overview of the \name architecture. Next, we describe the process of constructing MCPG. Then, we describe the detailed process of graph embedding, i.e., how to extract the node features and edge features of a graph. Finally, we explain the process of graph representation learning using NE-GCN.

\subsection{Overview}

Figure \ref{f1} presents the overall architecture of \name, consisting of three main steps. (1) Merging CPGs, which merges CPGs generated by buggy programs and fixed programs respectively into MCPG. MCPG is a joint graph structure that combines syntactic and semantic information about buggy programs and fixed programs, which provides a comprehensive view of patches. (2) graph embedding: which embeds the node attributes and edge attributes of MCPG into the feature vector. Node attributes contain the code snippet and node type for each MCPG node. Edge attributes contain the edge type and the edge version. (3) GCN learning, which is a well-designed graph convolutional neural network that performs message passing and aggregation of node feature vectors and edge feature vectors. 

\begin{figure*}[t]
\begin{center}
\includegraphics[width=0.8\textwidth]{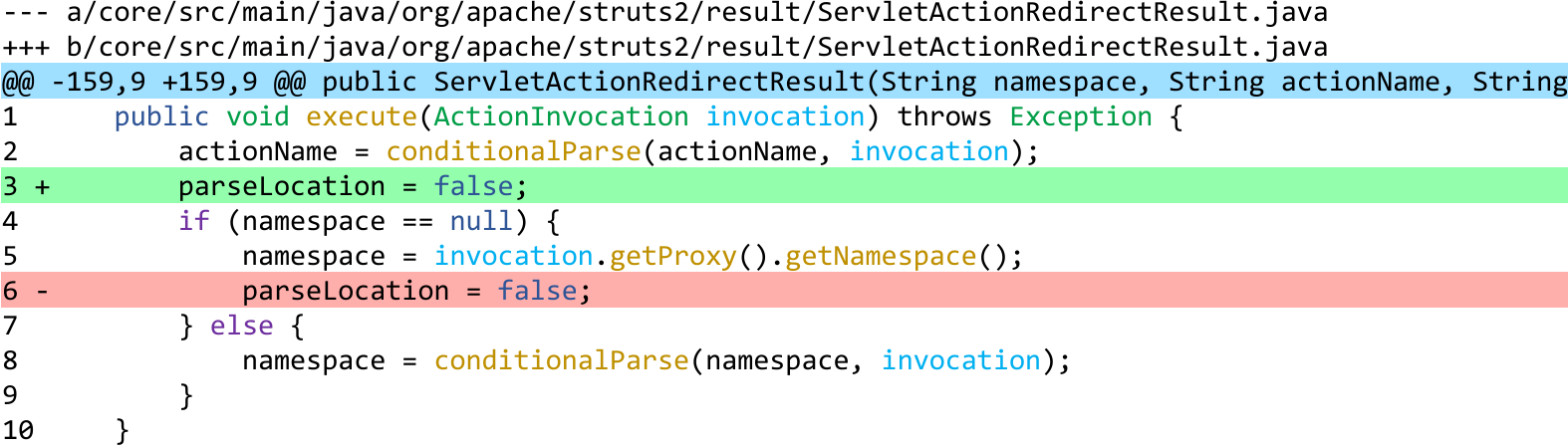}
\caption{A vulnerability fix in \textit{struts} project with its commit ID 6efaf90.}
\label{f2}
\end{center}
\end{figure*}

\subsection{Merging CPGs}
The merging CPGs section aims to merge graphical representations of buggy programs and fixed programs to obtain a more comprehensive and unified graph data structure. This part can be divided into four steps: source code acquisition, naming normalization, CPGs construction, and CPGs merging.

\subsubsection{\textbf{Source Code Acquisition}}We are concerned only with the files modified by the vulnerability fix patches. An example of a vulnerability fix patch in \textit{struts} project is shown in Figure \ref{f2}. Lines beginning with - - - and +++ identify the files modified by the patch, and lines beginning with - and + identify the lines of code removed and added, respectively. In order to encompass all critical code as much as possible, we need to obtain all the complete code files involved in the buggy program and the fixed program. We can utilize the commit ID to roll back the project to the defective versions, thus obtaining the relevant code files for both versions.

\subsubsection{\textbf{Naming Normalization}} We refer to related studies \cite{r25} and adopt a method of naming normalization  to alleviate the issue of vocabulary explosion and reduce the impact of specific variable names. This method maps programmer-designed function names, variable names, and strings one-to-one to specific symbolic names, such as fun1, var1, and str1. The uniform naming representation reduces the size of the vocabulary, thereby enhancing the model's generalization ability.

\subsubsection{\textbf{CPGs Construction}} For fixed programs and buggy programs, we use Joern \cite{r26} to parse. Each function in the code corresponds to a CPG. CPG is formulated as ${{G} }_{c p g}=\left({{V} }_{c p g}, {{E} }_{c p g}\right)$, where ${V}_{c p g}={V}_{a s t}$ and ${E}_{c p g}={E}_{a s t} \cup {E}_{c d g} \cup {E}_{d d g} \cup {E}_{c f g}$. $v$ contains the code and type information of the node and $e$ contains the type and version information of the edge. $V$ denotes the set of $v$ and $E$ denotes the set of $e$.

\subsubsection{\textbf{CPGs Merging}} For the constructed ${{G} }_{buggy}$ and ${{G} }_{fixed}$, merge is performed. The main purpose of merging ${{G} }_{buggy}$ and ${{G} }_{fixed}$ is to reflect the structural features of patches through a single graph form. This helps neural networks better understand the fixing behavior, thereby learning the unique patterns of vulnerability fixes. First, we determine the functions that have been modified in the patch and obtain the CPGs corresponding to the buggy version and the fixed version, respectively. To avoid conflicts in node IDs, we reassign the node IDs and update the corresponding node IDs in the edge set as well. Note that we start numbering the nodes of ${{G} }_{fixed}$ and then continue with those of ${{G} }_{buggy}$. If nodes remain unchanged, they retain their numbering from ${{G} }_{fixed}$. This means that ${{G} }_{fixed}$ and ${{G} }_{buggy}$ have partially overlapping nodes in the graph, which initially forms the merged graph ${{G} }_{merge}={{G} }_{buggy}\cup {{G} }_{fixed}$. Finally, determine the version of the node (fixed, buggy, both fixed and buggy) and the version of the edge (fixed, buggy), respectively, adding them to the elements of the node and the edge.

\begin{algorithm}
    \caption{slice ${{G}_{slice}}$}
    \label{alg1}
    \KwIn{${{G} }_{merge}=({{V} }_{merge},{{E} }_{merge}), {{V} }_{change}$}
    \KwOut{${{G}}_{slice}=({{V} }_{slice},{{E} }_{slice})$}
    
    initial ${{V}_{slice}, {E}_{slice}}=\left \{  \right \}, \left \{  \right \}  $ ;
    
    \tcp{step1: slicing on the graph}
    
    ${{T}}=\left \{ cdg,ddg,cfg  \right \}  $ ;
    
    \For{ $v$ in ${{V} }_{change}$}
    {
        
        \For{$e$ in ${{E} }_{merge}$}
        {

            \tcp{forward slicing}
            
            \If{$from\left\{ e\right\} =v$ \& $type(e)\in T$}
            {
                ${V}_{slice}={V}_{slice}\cup \left \{ v \right \}$, ${E}_{slice}={E}_{slice}\cup \left \{ e \right \}$\;
            }
            \tcp{backward slicing}
            
            \If{$to\left\{e\right\}=v$ \& $type(e)\in T$}
            {
                ${V}_{slice}={V}_{slice}\cup \left \{ v \right \}$, ${E}_{slice}={E}_{slice}\cup \left \{ e \right \}$\;
            }
        }
    }

    \tcp{step2: searching ast}

    \For{$e$ in ${{E} }_{merge}$}
    {
        
        \If{ $to\left \{ e \right \} \in {V}_{slice}$ $\mid$ $from\left \{ e \right \} \in {V}_{slice} $ }
        {
            \If{$type\left \{ e \right \} =ast$}
            {
                ${V}_{slice}={V}_{slice}\cup \left \{ to\left \{ e \right \} \right \}\cup \left \{ from\left \{ e \right \} \right \}$ \;
                ${E}_{slice}={E}_{slice}\cup \left \{ e \right \}$ \;
            }
        }
    }

    \Return ${{G}}_{slice}$\;

\end{algorithm}

In order to retain vulnerability-related code as much as possible and to reduce the impact of irrelevant code on the results, we perform program slicing on the graph. To comprehensively analyze the dependency and control relationships between various nodes, we perform both forward and backward slices on the graph. Backward slicing can help identify the source of some sensitive data in the program and understand the effect of variables, which is important for the security analysis of silent fixes. Forward slicing can assist in understanding the execution path of the program through control flow and data flow thus determining the scope of impact of code fragments.

Algorithm~\ref{alg1} describes the pseudo-code for slicing on graphs. Considering the well-merged CPGs, we first execute the slicing. Note that slicing is only performed in CDG, DDG, and CFG, where each node represents a slice statement. Forward slicing finds all nodes dependent on the given node and the corresponding edges, while backward slicing finds all nodes affecting the given node and the corresponding edges (lines 3-12). To reduce the noise, we select only one-hop neighbors of the slicing criterion during the slicing process. After performing the slicing, we retain the changed and sliced statements for all the nodes as well as the tracked edges. Subsequently, we continue to search the nodes and edges of the ast graph in the CPG. We traverse the slicing edge set obtained from the first step of the operation, find all edges whose start or end nodes are in the slicing point set, and add those edges and points that are not in the slicing node set to the slicing set (lines 13-20).

\begin{figure}[ht]
\begin{center}
\centering
\includegraphics[width=\linewidth]{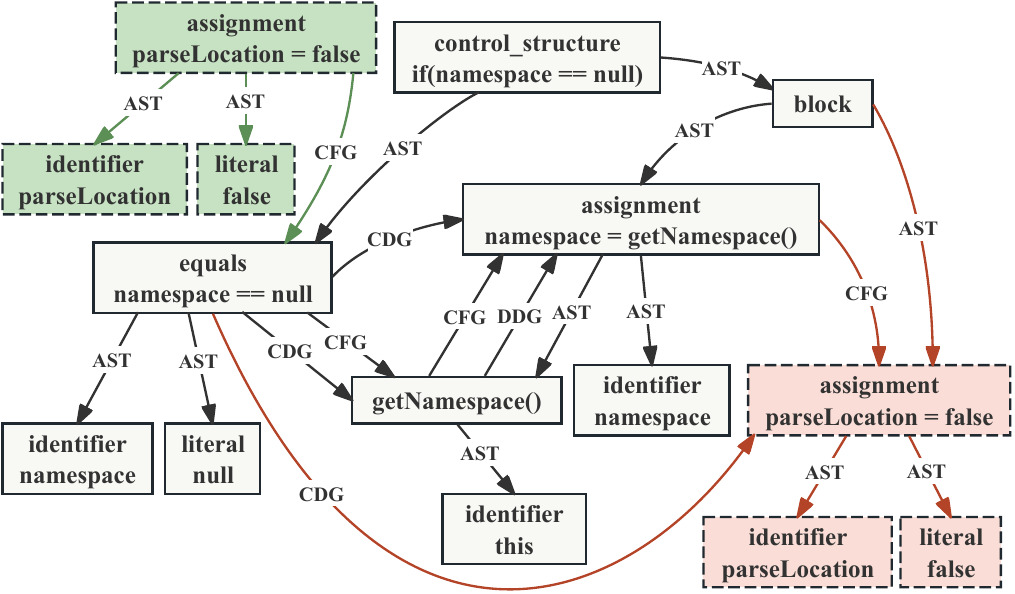}
\caption{MCPG of the patch in Figure \ref{f2}.}
\label{f3}
\end{center}
\end{figure}

After obtaining ${{G}}_{slice}$, we simplify the graph, and the related operations include removing some isolated nodes, removing nodes with empty code fragments, and so on. We define the simplified graph as MCPG. Figure \ref{f3} illustrates a complete MCPG constructed for the fix patch in Figure \ref{f2}, where the red/green color represents the deleted/added nodes and edges, and the node information includes code fragments and type information.

\subsection{Graph Embedding}
The graph embedding part is to parameterize the nodes and edges of each MCPG as an embedding (i.e., a numeric vector), aiming to capture the structural features of the graph in its entirety. Traditional graph embedding methods typically focus on learning low-dimensional representations of node codes, ignoring the feature information of edges. In our approach, we embed both codes and types of nodes as feature information. Also, the version and type information of edges are embedded as features of edges. The graph embedding is divided into node embedding and edge embedding.

\subsubsection{\textbf{Node Embedding}}Nodes are represented as a 5-tuple $(id,version,line,type,code)$ in the MCPG. To naturally preserve the complete information of nodes, we embed both the node types and code as features of the node. We traverse the node set to obtain node types and node codes as a training corpus. We choose Word2vec \cite{r27} as the vector representation model, which can efficiently represent rich semantic information. Types and codes need to be split to get the token sequences. After that, all the token sequences are mapped to a low-dimensional vector space using the Word2vec model. The feature information of a node $v\in V$ consists of the code $c$ and the type $t$. Token sequences corresponding to code $c$ and type $t$ are denoted by ${s}_{c}$, ${s}_{t}$ respectively. The embedding result $e$ of a node is made by concatenating ${e}_{c}$ and ${e}_{t}$, which denote the vector representation of ${s}_{c}$, ${s}_{t}$ resulting from Word2vec mapping. $e$ is defined as:
\begin{equation}
\label{eq1}
	e_{v} = concat(e_{s_{t}},e_{s_{c}}),
\end{equation}
where $concat(\cdot )$ denotes the operation of concatenating vectors. The code snippet is always present in the node information. However, if node type does not exist, then it is concatenated with a zero vector.

\subsubsection{\textbf{Edge Embedding}}
Edges are represented as a 4-tuple $({id}_{1},{id}_{2},version,type)$.
The version information of an edge consists of three types, exist only in buggy code, exist only in fixed code, and exist in both versions. The type information of an edge includes four kinds, namely the four graph representations of CPG, including AST, DDG, CDG, and CFG. In order to better capture the interaction information of nodes and to more carefully model the graph structure, we use 6-dimensional vectors for embedding the information of edges $(buggy, fixed, AST, CDG, DDG, CFG)$. The buggy bit and the fixed bit indicate the version of the edge, and the last four bits indicate the type of the edge.

\subsection{GCN learning}
We have successfully obtained effective embedding vectors for nodes and edges, and the subsequent task is to explore the characteristics of neighboring nodes. Inspired by the concept of message passing, we have designed a graph neural network model, referred to as NE-GCN, which passes and fuses the features of its neighbors along the edges. 

\subsubsection{\textbf{Message Passing}}
Researchers \cite{r29} \cite{r28} have amply demonstrated the powerful capabilities of message passing mechanisms in learning the structural characteristics of graphs. Traditional GCN only conveys and aggregates the feature information of nodes. Building on this foundation, we further explore the process of incorporating attribute features of edges into message passing. We propose an edge-aware message-passing framework that synthesizes node attributes and edge attributes in each iteration to enhance the model's ability to understand and represent graph structures. Within this framework, a node aggregates information from its neighboring nodes through a weighted aggregation function that also considers edge attributes in particular. To enable the model to adaptively capture the semantic significance of edge attributes, we introduce an edge feature transformation layer that employs a set of trainable parameters to optimize the representation of edge features and thus dynamically adjusts the weights of edge features in node feature updates.

Node feature matrix $X$ is of size $N\times D$, where $N$ represents the number of nodes in the graph, and $D$ denotes the dimensionality of each node's feature vector. In this matrix, each row $X_{i}$ corresponds to a feature vector $e_{v}$ of a particular node. 
Edge feature matrix $E$ then provides a description of the properties of edges in the graph, where each row corresponds to an edge and each column represents a feature of the edge.
The activation feature matrix $H$ has $H^{(0)}=X$ as its initial state and evolves to capture the complex relationships between nodes and the global structure of the graph as the message passing iterates.
The adjacency matrix $A$ detailedly describes the topological structure of the graph, with elements $A_{ij}$ indicating the connectivity between node $i$ and node $j$. 
Based on the above definition, the information propagated from neighboring nodes can be represented as:
\begin{equation}
\label{eq3}
	Z=\hat{D}^{-\frac{1}{2}} \hat{A} \hat{D}^{-\frac{1}{2}} H^{(l)} W^{(l)}_{n}+ EW^{(l)}_{e}
\end{equation}
where $\hat{A}=A+I$, $I$ is the identity matrix and $\hat{D}$ is the diagonal node degree matrix of $\hat{A}$. $\hat{D}^{-\frac{1}{2}} \hat{A} \hat{D}^{-\frac{1}{2}}$ is a degree normalization of $\hat{A}$. This operation is intended to balance the influence of different nodes during the message-passing process, particularly for those nodes with a high degree. $W^{(l)}_{n}$ and $W^{(l)}_{e}$ are the trainable weight matrices of the nodes and edges in layer $l$, respectively. 

Upon receiving information passed from neighboring nodes, we employ the idea of residual linking to integrate this multi-layer processed information with the original node features. This approach not only preserves the original features in the nodes, but also allows the model to capture more complex structures in each layer. We also choose $ReLU$ as the activation function, and the final $H^{(l+1)}$ is updated as:
\begin{equation}
\label{eq4}
	H^{(l+1)}=ReLU\left(Z+H^{(l)}\right )
\end{equation}

To further integrate feature information from multi-hop neighbors, we set up three layers of iterative propagation to update the node representation. The results of each iteration are concatenated and defined as follows:
\begin{equation}
\label{eq5}
H = concat(H^{(1)},H^{(2)},H^{(3)})
\end{equation}

Through this fine-grained feature integration and adaptive learning mechanism, the model's deep understanding of graph structure is facilitated.

\subsubsection{\textbf{Pooling and Classification}}
We handle the various tasks of vulnerability fix patches from the perspective of graph classification, and below we proceed to generate graph-level embeddings. At the pooling layer, we choose SAGPool \cite{r30} for graph pooling to simplify the structure of the graph. SAGPool utilizes the self-attention mechanism for graph pooling, which can take full account of the node features and the topology of the graph. In the readout layer, we combine max-pooling and average-pooling to extract the remarkable features and average features of the graph and concatenate them to get a compact graph representation. Finally, by applying multilayer perceptron and softmax layers to the graph representation, we can construct a framework for graph classification.

\section{Experiments Setup}

\subsection{Research Questions}
In this section, we focus on assessing the effectiveness of \name by answering the following research questions (RQs):

RQ1: How does \name perform in vulnerability fixes identification, compared with the state-of-the-art baselines?

RQ2: How effective is \name in classifying vulnerability types and vulnerability severity compared to the baselines?

RQ3: What is the impact of the design of different modules on the performance of \name?

\subsection{Dataset}
We manually constructed a balanced vulnerability fix dataset. The dataset incorporates two studies on vulnerability fixes, where the positive samples are derived from a widely used vulnerability dataset \cite{r32} and the negative samples are derived from a security patch dataset \cite{r33}.
The dataset \cite{r32} covers 624 publicly documented vulnerabilities in 205 open source Java projects and the corresponding 1,282 fix commits. By crawling these fix commits from GitHub and performing thorough data cleaning, we excluded invalid commits such as those that only modified non-Java files or those whose code changes lacked practical significance. This process allowed us to refine a sample of 1,068 high-quality vulnerability fixes.
The dataset \cite{r33} contains 110,000 non-security-related commits. We first filtered out commits originating from the Java project. The commits were then scrutinized to exclude keywords related to security issues. Finally, we selected commits that originated from projects with high stars on GitHub to ensure the quality and relevance of the sample. With this rigorous series of screening steps, we obtained a sample of 1,183 non-vulnerability fixes. 
We ended up with a balanced and representative vulnerability fix dataset, thus laying a solid data foundation for subsequent research in the field of vulnerability fix.

To enrich our vulnerability fix dataset and provide deeper insights for the study, we classified and labeled the vulnerability fixes in the dataset in detail, based on the CWE category and CVSS severity to which they belong.
We queried the NVD to obtain the CWE category and CVSS score corresponding to the CVE ID of each vulnerability. Given the breadth and diversity of the CWE categories, we appropriately grouped those with less frequent occurrences to facilitate a more focused and efficient analysis.
In terms of CVSS scores, due to a certain lag in manual labeling, some vulnerabilities failed to obtain CVSS scores. Meanwhile, we referred to the classification method in CVSS 3.0 to classify the severity of vulnerabilities into four levels. This classification not only simplifies the scoring system but also makes the dataset more intuitive and easy to handle when analyzing and applying it. 
As a result, our dataset provides detailed information about vulnerability fixes to better understand and analyze the potential impact of vulnerabilities.

\subsection{Implementation Details}
In the merging CPGs phase, we use Python scripts for naming normalization and joern for parsing Java code into CPGs. Then the process of building MCPG is also developed by Python code.
In the graph embedding phase, we initialize the embedding of each node via Word2vec. Given the best performance, we set the embedding size of Word2vec to 64, the embedding size of node code snippets to 12, and the embedding size of node types to 4.
In the GCN learning phase, we implemented NE-GCN using the deep learning library PyTorch with PyTorch Geometric as the graph neural network framework. We use a ratio of 4:1 to split the train sets and test sets. In NE-GCN, the number of layers for message passing is set to 3, the batch size is set to 32, and the epoch is set to 50. We use the Adamax optimization algorithm for training the model and the exponential decay strategy for adjusting the learning rate. The learning rate is $1\cdot 10^{-3}$, weight decay is $5\cdot 10^{-4}$, and gamma is 0.8. To mitigate the overfitting phenomenon, we implement a technique called dropout, setting the dropout rate to 0.5. This strategy prevents the model from over-relying on the training data and enhances the generalization ability of the model. For assessing multi-classification tasks, we adopted the macro-average method. This method ensures that each category has equal weight in the evaluation. To verify the stability and reliability of the results, all experiments were repeated independently five times, with the outcomes being largely consistent. Ultimately, we chose the data from the last experiment as our reported final results.

All experiments are conducted on a server with 12 cores Intel Xeon Silver 4214R CPUs @ 2.40GHz, an RTX 3080Ti GPU with 12GB of RAM, and 90GB physical memory.

\subsection{Baselines}
In order to validate the effectiveness of \name in handling tasks related to vulnerability remediation patching, we have carefully selected three state-of-the-art methods in the field of silent vulnerability fix and three neural network models that have been widely used in the field of Natural Language Processing (NLP).

\textbf{Transformer} \cite{r2}:  This is a novel neural network architecture that relies entirely on attention mechanisms for sequence transduction tasks, eliminating the need for recurrent or convolutional layers. The Transformer model utilizes multi-head self-attention and positional encodings to process input sequences in parallel.

\textbf{TextCNN} \cite{r1}: This paper uses Convolutional Neural Networks for sentence-level classification tasks, in which, Convolutional Neural Networks are used to capture semantic features based on pre-trained word vectors.

\textbf{Bi-LSTM} \cite{r3}: The paper introduces a bidirectional recurrent neural network (BRNN) that can be trained in both positive and negative time directions, allowing it to utilize all available past and future information for prediction without the need for a preset delay

\textbf{PatchRNN} \cite{r5}: This method combines textual analysis of commit messages with syntactic and semantic analysis of source code changes to identify security patches in open source software. It employs a twin Recurrent Neural Network to process the code differences.

\textbf{SPI} \cite{r6}: This work utilizes two neural networks, one to analyze commit messages and another to learn from code revisions, combining their insights to accurately detect security patches.

\textbf{GraphSPD} \cite{r4}: This is a graph-based security patch detection system that merges pre-patch and post-patch code property graphs. Its model utilizes multi-attributed graph convolution to detect if a patch is security-related directly from its graph-structured representation.

\subsection{Evaluation Metrics}
We use Accuracy, Precision, Recall, F1-score, and FPR to evaluate the effectiveness of vulnerability fix identification. For the two multi-classification tasks, CWE category classification and vulnerability severity classification, we use Accuracy, F1-score, and MCC to evaluate the performance.

\textbf{Accuracy} is a measure of how often a classification model is correct in predicting the outcomes for both positive and negative classes, expressed as a ratio of the correctly classified instances to the total number of instances.

\textbf{Precision}  indicates the proportion of actual positives that were identified correctly out of all instances the model classified as positive, highlighting the model's ability to avoid false positives.

\textbf{Recall}, also known as sensitivity, measures the proportion of actual positives that were detected by the model out of all the actual positives, focusing on the model's ability to capture all relevant instances.

\textbf{F1-score} is the harmonic mean of precision and recall, providing a single metric that balances both false positives and false negatives, offering a more comprehensive view of the model's performance.

\textbf{False Positive Rate} is the ratio of negative classes that were incorrectly classified as positive by the model to the total number of actual negative classes, highlighting the model's propensity to generate false alarms.

\textbf{MCC} handles category imbalance well when evaluating classification models. It ranges from -1 to 1, where 1 indicates perfect categorization, 0 indicates random categorization and negative values indicate worse than random categorization.

\section{Results}
\subsection{\textbf{RQ1: Performance on Vulnerability Fixes Identification}}

\subsubsection{\textbf{Settings}}To evaluate how \name performs on vulnerability fix identification, we use a constructed dataset of balanced positive and negative samples, including 1068 vulnerability fix patches and 1183 non-vulnerability fix patches.
We compare \name to six neural network models, including three approaches common to the natural language processing field and three solutions for the vulnerability fix identification task.

\renewcommand{\thetable}{\arabic{table}}

% Table generated by Excel2LaTeX from sheet 'Sheet6'
% \begin{table}[htbp]
%   \centering
%   \caption{Results of Vulnerability Fix Identification}
%     \setlength{\tabcolsep}{2.8mm}
%     \begin{tabular}{c|cccc}
%     \toprule
%     \textbf{Method} & \textbf{Accuracy} & \textbf{Precision} & \textbf{Recall} & \textbf{F1-score} \\
%     \midrule
%     \midrule
%     \textbf{Transformer} & 74.28 & 74.19 & 74.36 & 74.20 \\
%     \textbf{TextCNN} & 77.38 & 77.26 & 77.06 & 77.14 \\
%     \textbf{Bi-LSTM} & 78.73 & 80.10 & 72.37 & 76.04 \\
%     \textbf{GraphSPD} & 64.75 & 63.72 & 68.50 & 66.02 \\
%     \textbf{PatchRNN} & 81.56 & 79.91 & 80.62 & 80.30 \\
%     \textbf{SPI} & 82.04 & 81.92 & 82.15 & 81.97 \\
%     \midrule
%     \textbf{\name} & \textbf{89.14} & \textbf{88.73} & \textbf{88.32} & \textbf{88.52} \\
%     \bottomrule
%     \end{tabular}%
%   \label{t1}%
% \end{table}%

\subsubsection{\textbf{Results}}Table \ref{t1} demonstrates the performance comparison of the generic approaches and the vulnerability fix identification approaches with \name on the constructed dataset. Overall, using Merge-CPGs as the graphical representation of patches and NE-GCN as the graph feature extraction model, \name outperforms the six baseline methods on four metrics. Compared to the widely-used generic approaches, \name improves 10.41\% (Bi-LSTM) to 14.86\% (Transformer) in accuracy, 8.63\% (Bi-LSTM) to 14.54\% (Transformer) in precision, 11.26\% (TextCNN) to 15.95\% (Bi-LSTM) in Recall, 11.38\% (TextCNN) to 14.32\% (Transformer) in F1-score, and decreases by 6.09\% (Bi-LSTM) to 13.21\% (Transformer) in FPR. Compared to specialized vulnerability fix approaches, \name improves 7.1\% (SPI) to 24.39\% (GraphSPD) in accuracy, 6.81\% (SPI) to 25.01\% (GraphSPD) in precision, 6.17\% (SPI) to 19.82\% (GraphSPD) in recall, 6.55\% (SPI) to 22.50\% (GraphSPD) in F1-score, and decreases by 6.23\% (SPI) to 25.07\% (GraphSPD). In addition, \name achieved the best overall performance with 89.14\% in accuracy, 88.73\% in precision, 88.32\% in recall, 88.52\% in F1-score, and 10.14\% in FPR.

% Table generated by Excel2LaTeX from sheet 'Sheet6'
\begin{table}[htbp]
  \centering
  \caption{Results of Vulnerability Fix Identification}
    \setlength{\tabcolsep}{0.8mm}
    \begin{tabular}{c|ccccc}
    \toprule
    \textbf{Method} & \textbf{Accuracy/\%} & \textbf{Precision/\%} & \textbf{Recall/\%} & \textbf{F1-score/\%} & \textbf{FPR/\%} \\
    \midrule
    \midrule
    \textbf{Transformer} & 74.28 & 74.19 & 74.36 & 74.20 & 23.35 \\
    \textbf{TextCNN} & 77.38 & 77.26 & 77.06 & 77.14 & 20.48 \\
    \textbf{Bi-LSTM} & 78.73 & 80.10 & 72.37 & 76.04 & 16.23 \\
    \textbf{GraphSPD} & 64.75 & 63.72 & 68.50 & 66.02 & 35.21 \\
    \textbf{PatchRNN} & 81.56 & 79.91 & 80.62 & 80.30 & 18.30 \\
    \textbf{SPI} & 82.04 & 81.92 & 82.15 & 81.97 & 16.37 \\
    \midrule
    \textbf{\name} & \textbf{89.14} & \textbf{88.73} & \textbf{88.32} & \textbf{88.52} & \textbf{10.14} \\
    \bottomrule
    \end{tabular}%
  \label{t1}%
\end{table}%

\subsubsection{\textbf{Analysis}}It is clear from the experimental results that our approach achieves optimal performance in the vulnerability fix identification task. It illustrates the strong understanding of \name towards fixing patches, which is able to accurately recognize the difference between vulnerability fixes and non-vulnerability fixes. Our approach also makes a great trade-off between sensitivity (covering as many real vulnerability fixes as possible) and reliability (identifying vulnerability fixes as accurately as possible). In addition, we notice that GraphSPD, a graph-based method, does not perform well in the experiments. We attribute this to the fact that its node embedding dimension is only 20 dimensions, resulting in its limited ability to capture vulnerability characterization information.

\name uses MCPG as patch representations, which greatly enriches the structural information that can be captured by the model.
By conducting slicing and selection operations at the graph level, \name skillfully integrates the CPGs of buggy programs and fixed programs. This process not only accurately preserves the context nodes that are closely related to defects, but also effectively reduces the negative impact of noise without losing information integrity.
\name employs a well-designed GCN architecture, which enables the model to comprehensively reveal the implicit features of vulnerabilities by aggregating and passing the edge attributes with the node attributes.

\begin{tcolorbox}[colframe=black,colback=gray!12,boxrule=1pt,left=3.5mm,right=3.5mm,top=1.5mm,bottom=1.5mm]
\textbf{Answer to RQ1:} \name outperforms three generalized deep learning methods and three state-of-the-art vulnerability fix identification methods on all evaluation metrics, which shows that \name can pick up real vulnerability features in patches.
\end{tcolorbox}

\subsection{\textbf{RQ2: Performance on Vulnerability Fixes assessment}}

\begin{figure}[ht]
\begin{center}
\centering
\includegraphics[width=\linewidth]{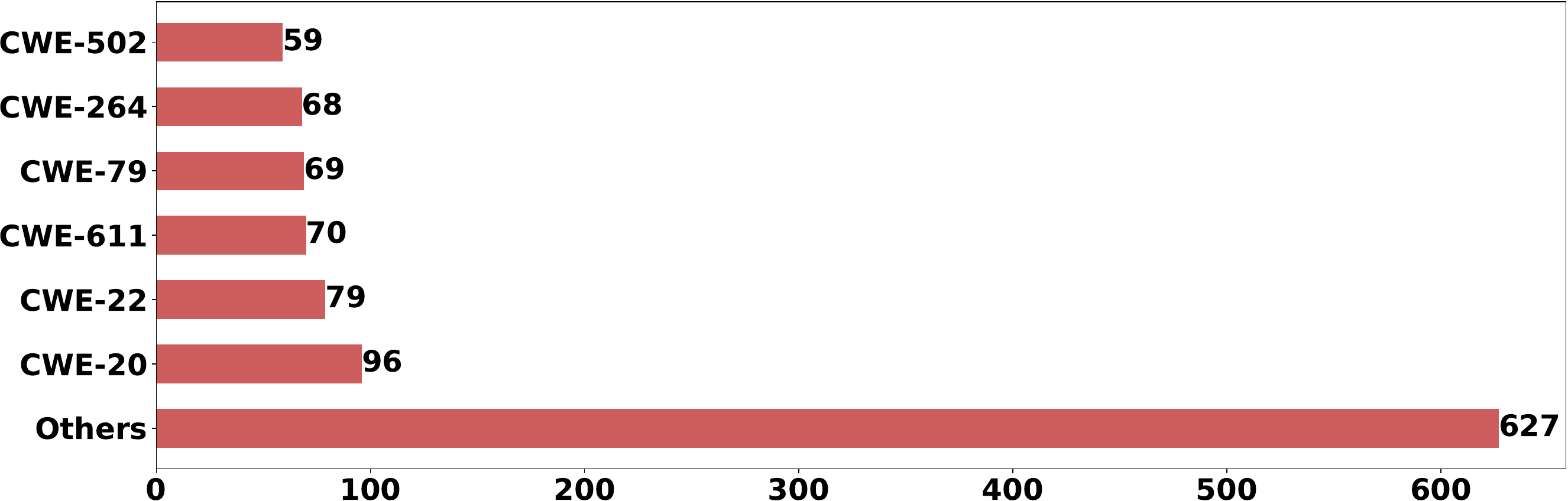}
\caption{Distribution of Vulnerable CWE Types.}
\label{f6}
\end{center}
\end{figure}

\subsubsection{\textbf{Settings}}To evaluate the performance of the model in vulnerability fix assessment, we label the vulnerability fixes separately. For the CWE classification task, we use a dataset consisting of 1068 vulnerability fix patches. As shown in Figure \ref{f6}, the labels include the 6 most common CWE categories in the dataset and other categories. For the severity classification task, we use a dataset consisting of 755 vulnerability fix patches with CVSS scores. The severity of vulnerabilities is classified as low, medium, high, or critical according to the CVSS 3.0 classification method.
Due to the lack of relevant open source methods, we only use three widely-recognized deep-learning methods as a baseline.

\begin{table}[htbp]
  \centering
  \caption{Results of Vulnerability CWE Classification}
    \setlength{\tabcolsep}{2mm}
    \begin{tabular}{c|ccccc}
    \toprule
    \textbf{Method} & \textbf{Accurary} & \textbf{Precision} & \textbf{Recall} & \textbf{F1-score} & \textbf{MCC} \\
    \midrule
    \midrule
    \textbf{Transformer} & 72.69 & 90.88 & 46.59 & 56.07 & 0.530 \\
    \textbf{TextCNN} & 66.20 & 53.54 & 45.01 & 45.86 & 0.420 \\
    \textbf{Bi-LSTM} & 68.52 & 57.68 & 49.75 & 51.61 & 0.480 \\
    \midrule
    \textbf{\name} & \textbf{79.63} & \textbf{81.00} & \textbf{61.90} & \textbf{68.37} & \textbf{0.657} \\
    \bottomrule
    \end{tabular}%
  \label{t3}%
\end{table}%

\subsubsection{\textbf{Results}}Table \ref{t3} demonstrates the performance of \name compared to three widely used models for the CWE category classification task. It can be seen that \name outperforms the three baseline methods on three metrics. Compared to the widely used generic approaches, \name improves accuracy by 4.62\% (Transformer) to 10.18\% (TextCNN), F1-score by 12.57\% (Transformer) to 25.64\% (TextCNN), and MCC by 0.102 (Transformer) to 0.188 ( TextCNN) MCC. \name achieved the overall best performance with 77.31\% in accuracy, 67.93\% in F1-score and 0.625 in MCC.

\begin{table}[htbp]
  \centering
  \caption{Results of Vulnerability Severity Classification}
    \setlength{\tabcolsep}{2mm}
    \begin{tabular}{c|ccccc}
    \toprule
    \textbf{Method} & \textbf{Accurary} & \textbf{Precision} & \textbf{Recall} & \textbf{F1-score} & \textbf{MCC} \\
    \midrule
    \midrule
    \textbf{Transformer} & 66.89 & 84.11 & 67.24 & 69.05 & 0.493 \\
    \textbf{TextCNN} & 64.86 & 47.88 & 46.79 & 46.54 & 0.437 \\
    \textbf{Bi-LSTM} & 71.62 & 59.93 & 59.89 & 58.35 & 0.556 \\
    \midrule
    \textbf{\name} & \textbf{79.73} & \textbf{84.04} & \textbf{75.07} & \textbf{78.57} & \textbf{0.681} \\
    \bottomrule
    \end{tabular}%
  \label{t4}%
\end{table}%

Table \ref{t4} shows how \name's performance compares to three widely recognized natural language processing methods in the vulnerability severity classification task. Overall, \name performs far better than the three baseline methods on all metrics. Compared to the baseline, \name improves the accuracy by 8.78\% (Bi-LSTM) to 14.19\% (TextCNN), the F1-score by 14.98\% (Bi-LSTM) to 19.57\% (TextCNN), and the MCC by 0.142 (Bi-LSTM) to 0.213 (TextCNN). \name achieves the best results with 77.31\% in accuracy, 67.93\% in F1-score and 0.625 in MCC.

\subsubsection{\textbf{Analysis}}\name achieves excellent performance on the two vulnerability-related multi-classification tasks. We attribute this to its advanced architectural design and feature extraction capabilities. \name utilizes a well-designed GCN architecture to effectively integrate multi-dimensional information about vulnerability patches, including code structure, syntax information, semantic content, and key features associated with vulnerabilities. 

The improvement of \name in F1-score and MCC further confirms its strength in dealing with class imbalance. These enhancements show that \name is not only able to correctly classify vulnerability fixes, but also maintains a fine balance between multiple samples, which is particularly important for practical vulnerability fix assessment.

\begin{tcolorbox}[colframe=black,colback=gray!12,boxrule=1pt,left=3.5mm,right=3.5mm,top=1.5mm,bottom=1.5mm]
\textbf{Answer to RQ2:} GRAPE demonstrates a significant advantage over the baseline in both F1-score and MCC, which indicates that it can effectively capture the diverse characteristics of vulnerabilities fix.
\end{tcolorbox}

\subsection{\textbf{RQ3: Impact of \name modules}}

\begin{figure}[htbp]
\begin{minipage}[]{0.32\linewidth}
\centering
\includegraphics[width=\linewidth]{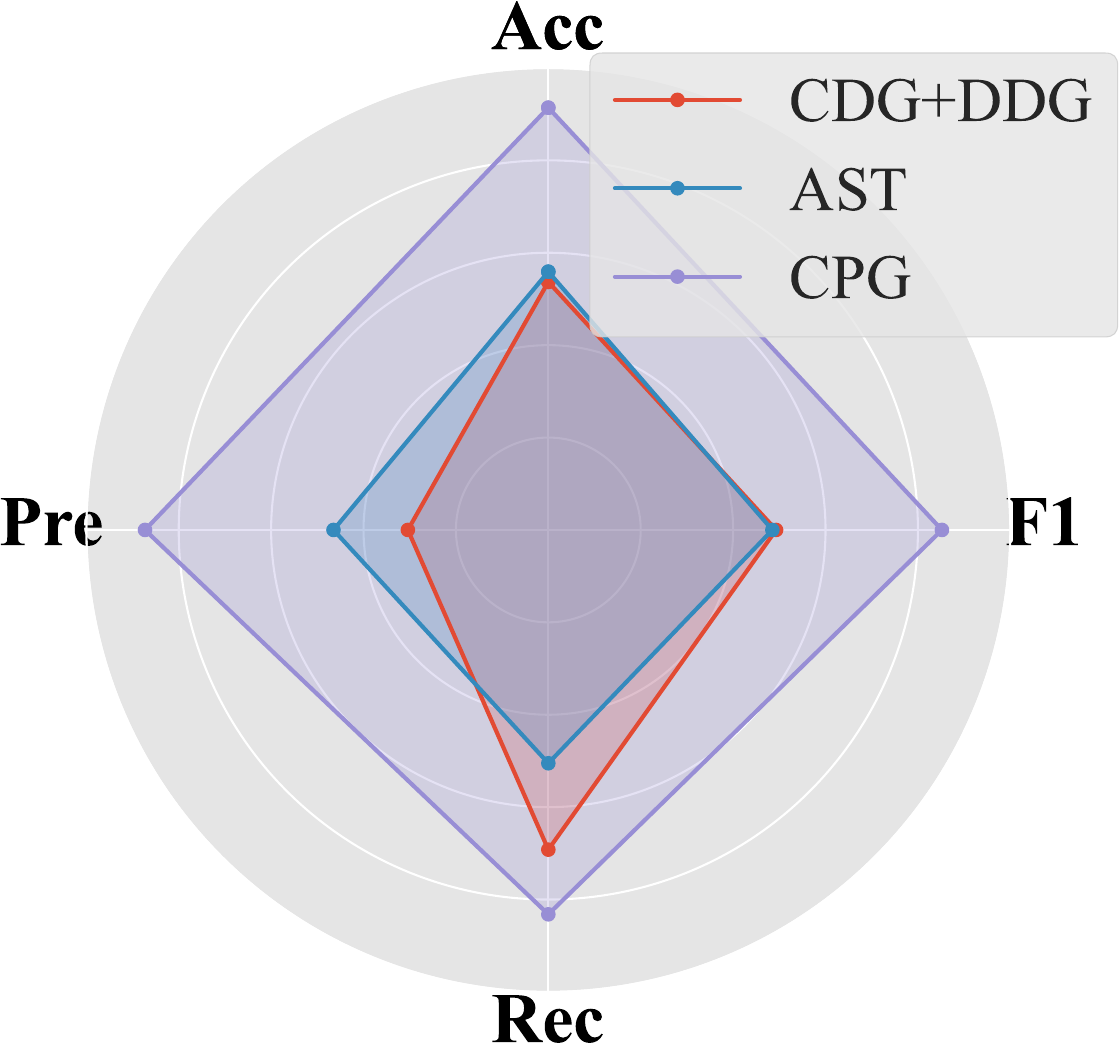} 
\caption*{(\textbf{a})VFI}
\end{minipage}
\begin{minipage}[]{0.32\linewidth}
\centering
\includegraphics[width=\linewidth]{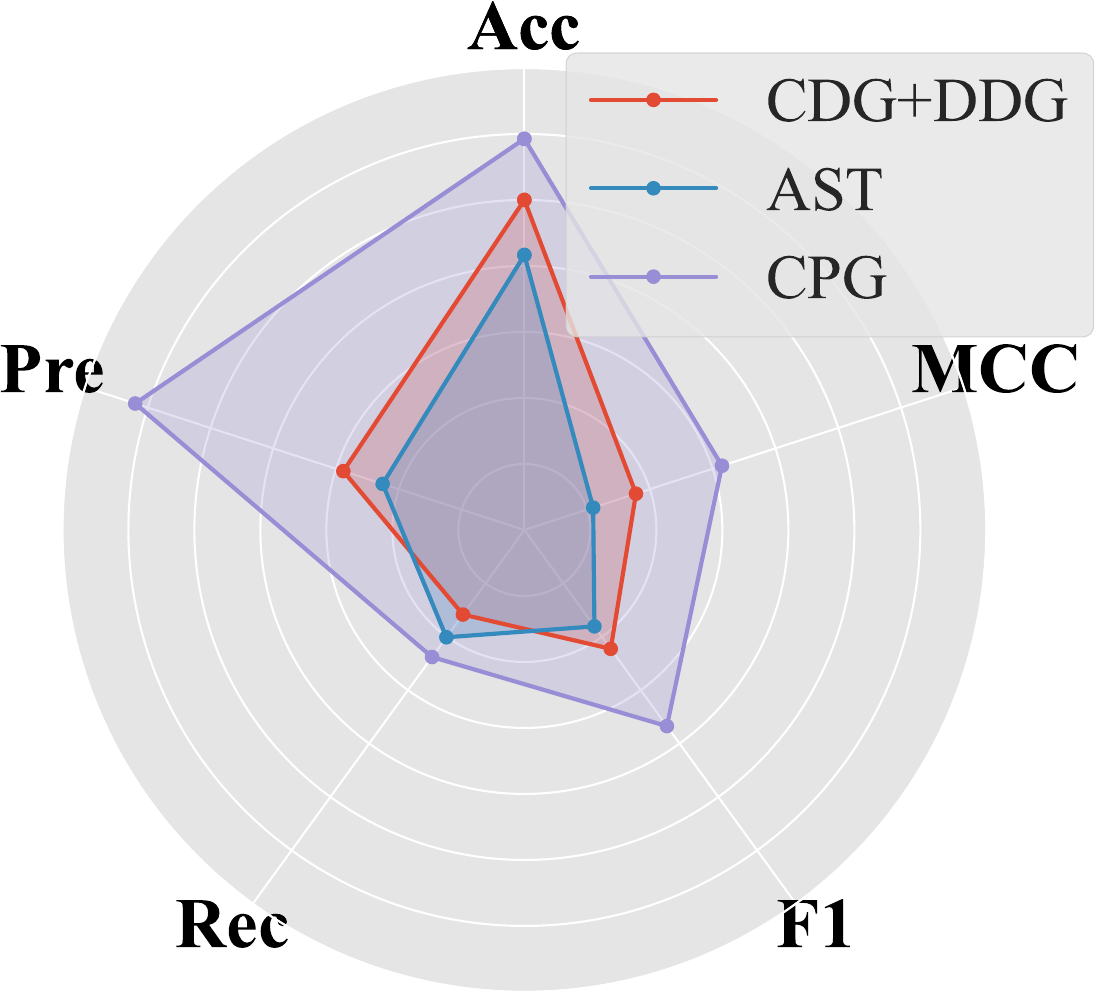}
\caption*{(\textbf{b})VTC}
\end{minipage}
\begin{minipage}[]{0.32\linewidth}
\centering
\includegraphics[width=\linewidth]{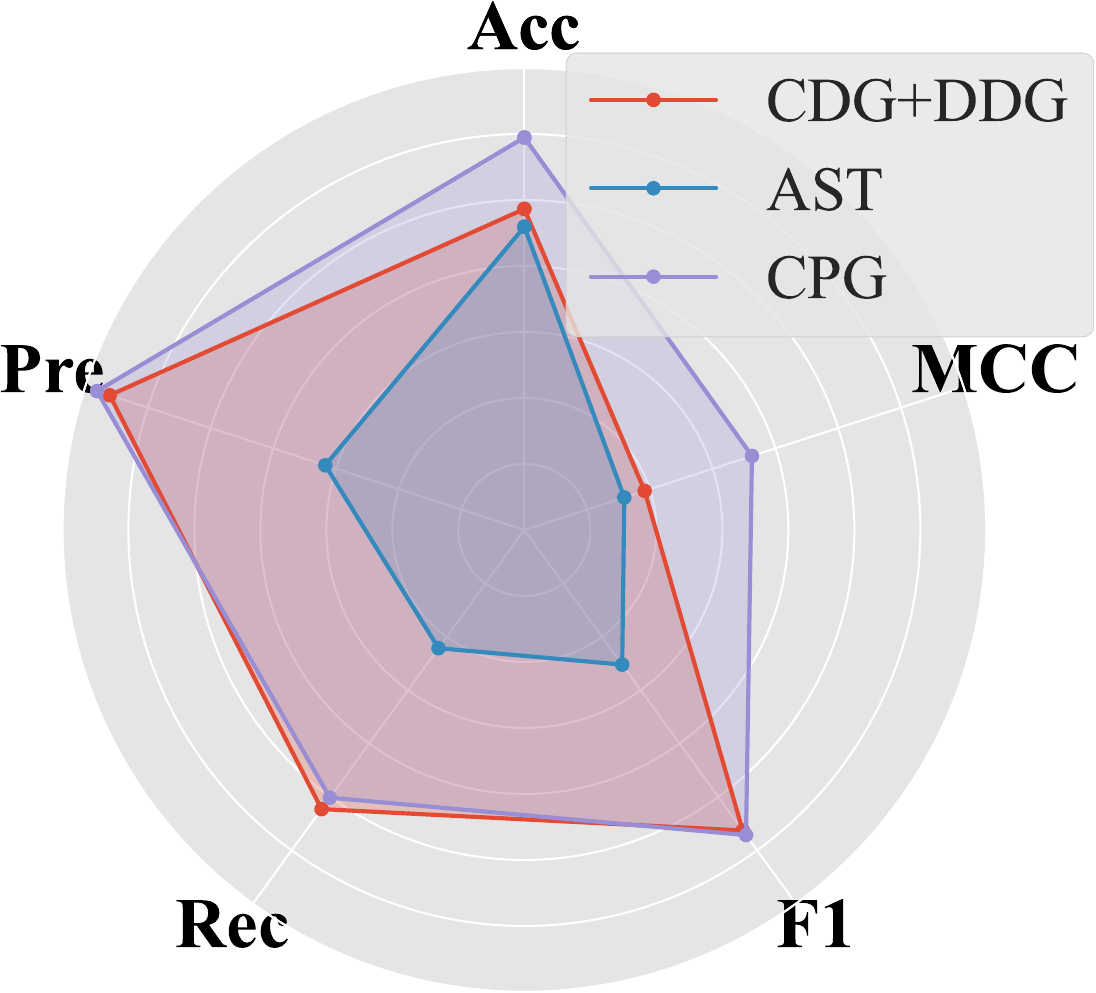} 
\caption*{(\textbf{c})VSC}
\end{minipage}
\caption{Performance of Different Graph Data Structures.}
\label{f5}
\end{figure}

\subsubsection{\textbf{Effect of choosing CPG as a data structure}}
We choose CPG as the graph data structure for the source code. To evaluate the effectiveness of CPG in patch representation, we compare its performance with AST and DDG+CDG. The three radar graphs in Figure \ref{f5} compare the performance of AST, DDG+CDG, and CPG in Vulnerability Fix Identification (VFI), vulnerability types classification (VTC), and Vulnerability Severity Classification (VSC). The experimental results clearly show that the best performance is obtained by using CPG. This result is expected because CPG not only captures the syntactic structure of the program, but also incorporates control dependencies and data dependencies between nodes. It provides a comprehensive view of the code structure and behavior. Compared to AST, the performance of DDG+CDG also presents an advantage, which indicates that inter-code dependencies contribute significantly to code modeling.

\begin{figure}[htbp]
\begin{minipage}[]{0.49\linewidth}
\centering
\includegraphics[width=\linewidth]{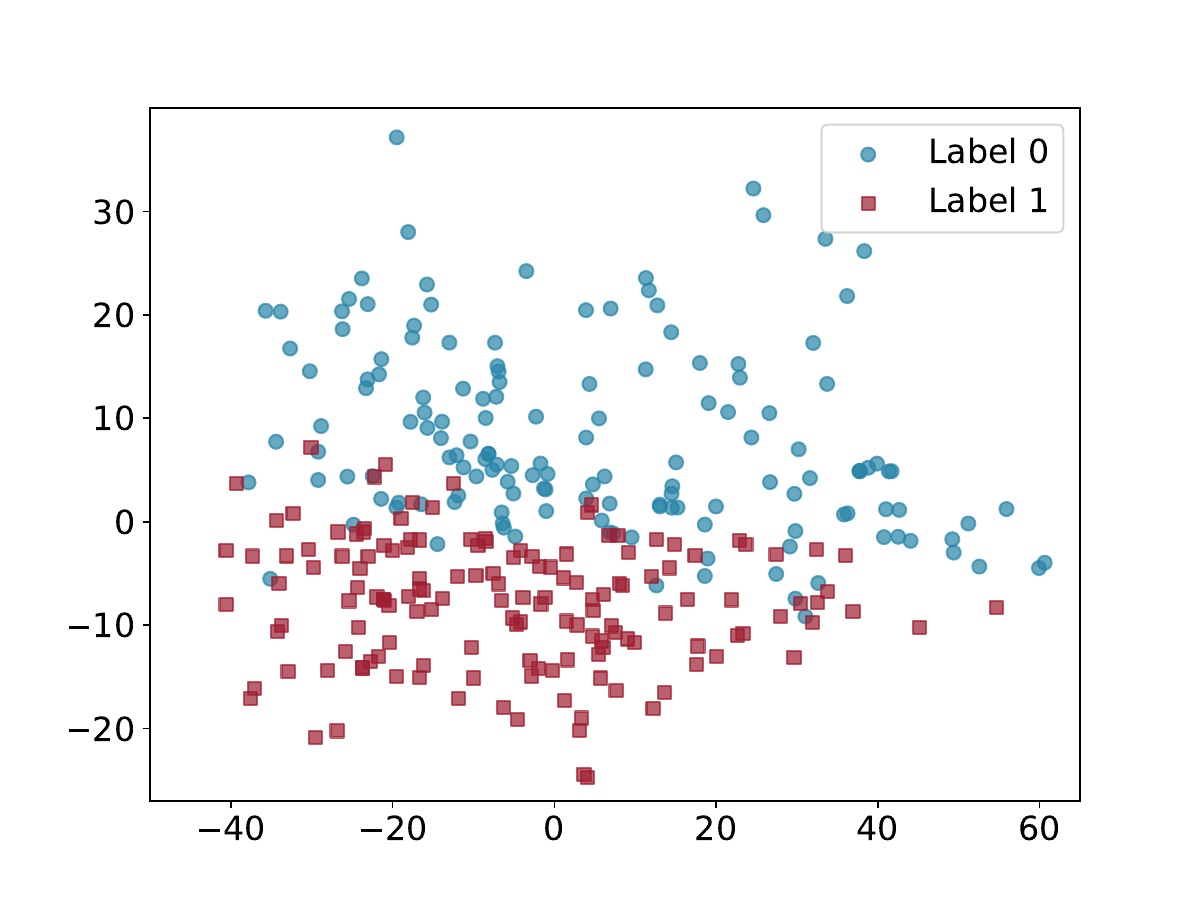}
\caption*{(\textbf{a}) \name(with)}
\end{minipage}
\begin{minipage}[]{0.49\linewidth}
\centering
\includegraphics[width=\linewidth]{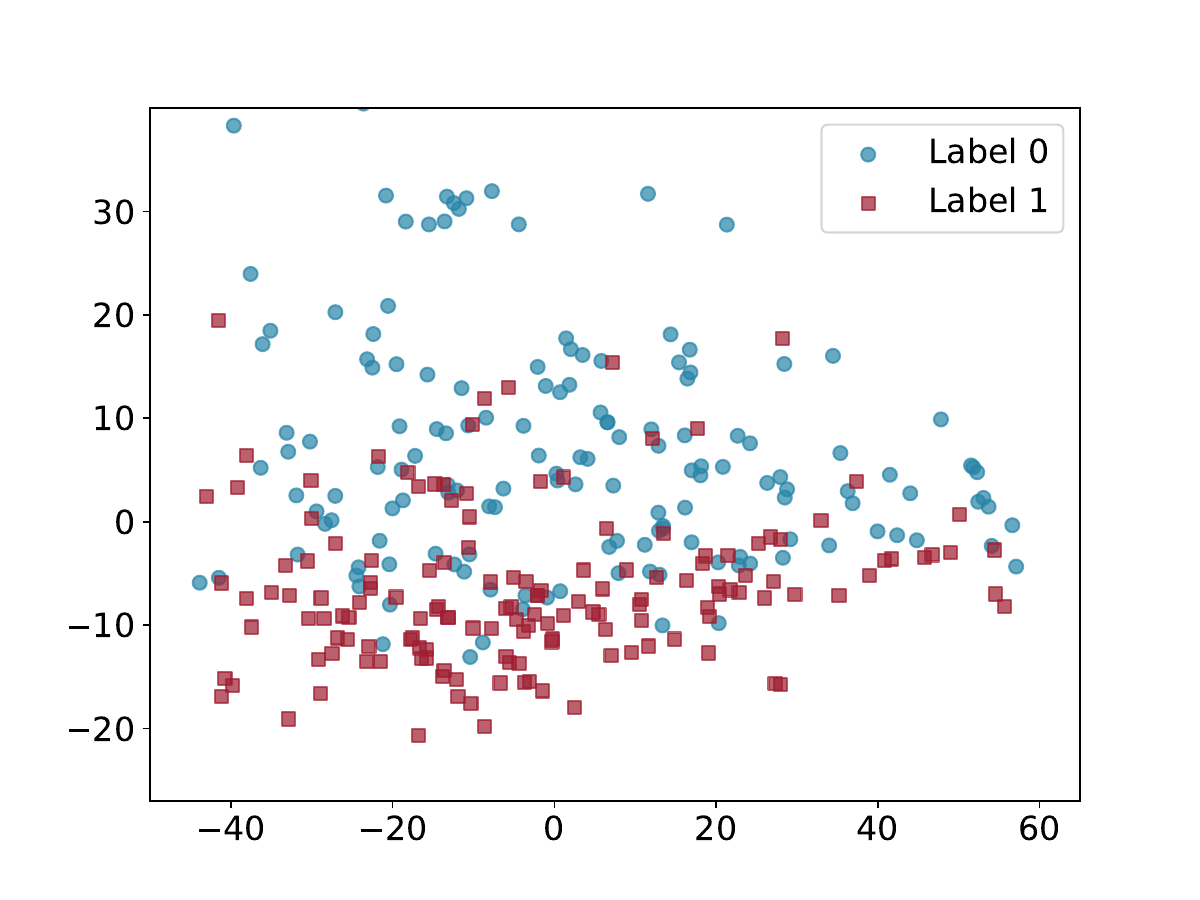} 
\caption*{(\textbf{b}) \name(without)}
\end{minipage}
\caption{Visualization of \name learned feature representation with/without graph merging module}
\label{f4}
\end{figure}

\subsubsection{\textbf{Effect of merging CPGs module}}After converting the buggy program and the fixed program into CPGs, we conduct a series of slicing and pruning operations to merge the graphs. To evaluate the effectiveness of this module, we employ PCA \cite{r58}, \cite{r59} techniques to visualize the learned patch representations. Focusing on the vulnerability fix identification task, we compare the visualization results obtained by using the graph merging module with those obtained by simple merging operations. In Figure \ref{f4}, Label 1 indicates a vulnerability fix, and Label 0 indicates a non-vulnerability fix. As depicted in Figure 6a, vulnerability fix patches and non-vulnerability fix patches are distinctly discernible in 2D space. Conversely, the experimental results without slicing and pruning operations, as illustrated in Figure 6b, exhibit greater overlap between the two groups. This discrepancy suggests that our proposed merging module is able to represent the graph structure more efficiently in the vulnerability fix identification task, thus enhancing the differentiation between vulnerability fix patches and non-vulnerability fix patches.
% Table generated by Excel2LaTeX from sheet 'Sheet7'
\begin{table}[htbp]
  \centering
  \caption{Performance of Different Node Embedding}
    \setlength{\tabcolsep}{0.4mm}
    \begin{tabular}{c|c|ccccc}
    \toprule
    \multirow{2}[4]{*}{\textbf{Task}} & \multirow{2}[4]{*}{\textbf{Embedding}} & \multicolumn{5}{c}{\textbf{Metric}} \\
\cmidrule{3-7}          &       & \textbf{Accuracy} & \textbf{Precision} & \textbf{Recall} & \textbf{F1-score} & \textbf{MCC} \\
    \midrule
    \midrule
    \textbf{Vulnerability Fix } & \textbf{Code} & 85.81 & 84.09 & 86.45 & 85.25 & - \\
    \textbf{Identification} & \textbf{Code+Type} & \textbf{89.14} & \textbf{88.73} & \textbf{88.32} & \textbf{88.52} & - \\
    \midrule
    \textbf{CWE} & \textbf{Code} & 75.93 & 80.15 & 59.27 & 64.14 & 0.591 \\
    \textbf{Classification} & \textbf{Code+Type} & \textbf{79.63} & \textbf{81.00} & \textbf{61.90} & \textbf{68.37} & \textbf{0.657} \\
    \midrule
    \textbf{Severity} & \textbf{Code} & 76.35 & 82.26 & 64.87 & 69.28 & 0.626 \\
    \textbf{Classification} & \textbf{Code+Type} & \textbf{79.73} & \textbf{84.04} & \textbf{75.07} & \textbf{78.57} & \textbf{0.681} \\
    \bottomrule
    \end{tabular}%
  \label{t6}%
\end{table}%

% Table generated by Excel2LaTeX from sheet 'Sheet4'
\begin{table*}[htbp]
  \centering
  \caption{Performance of Different GNNs on Vulnerability Fix Identification, CWE Classification, and Severity Classification}
  \setlength{\tabcolsep}{1.1mm}
    \begin{tabular}{c|cccc|ccccc|ccccc}
    \toprule
    \multirow{2}[4]{*}{\textbf{Method}} & \multicolumn{4}{c|}{\textbf{Vulnerability Fix Identification}} & \multicolumn{5}{c|}{\textbf{CWE Classification}} & \multicolumn{5}{c}{\textbf{Severity Classification}} \\
\cmidrule{2-15}          & \textbf{Accurary} & \textbf{Precision} & \textbf{Recall} & \textbf{F1-score} & \textbf{Accurary} & \textbf{Precision} & \textbf{Recall} & \textbf{F1-score} & \textbf{MCC} & \textbf{Accurary} & \textbf{Precision} & \textbf{Recall} & \textbf{F1-score} & \textbf{MCC} \\
    \midrule
    \midrule
    \textbf{GAT} & 83.59 & 82.11 & 83.64 & 82.87 & 71.30 & 65.42 & 57.73 & 60.38 & 0.530 & 69.59 & 77.67 & 67.95 & 70.31 & 0.544 \\
    \textbf{GCN} & 84.70 & 84.69 & 82.71 & 83.69 & 68.98 & 66.77 & 59.69 & 59.61 & 0.526 & 72.97 & 82.50 & 60.14 & 65.83 & 0.572 \\
    \textbf{GGNN} & 84.26 & 84.88 & 81.31 & 83.05 & 76.85 & 77.48 & \textbf{62.62} & 68.13 & 0.608 & 78.38 & 82.66 & 67.30 & 70.58 & 0.665 \\
    \midrule
    \textbf{\name} & \textbf{89.14} & \textbf{88.73} & \textbf{88.32} & \textbf{88.52} & \textbf{79.63} & \textbf{81.00} & 61.90 & \textbf{68.37} & \textbf{0.657} & \textbf{79.73} & \textbf{84.04} & \textbf{75.07} & \textbf{78.57} & \textbf{0.681} \\
    \bottomrule
    \end{tabular}%
  \label{t5}%
\end{table*}%

\subsubsection{\textbf{Effect of graph embedding modules}}In the graph embedding phase, \name initializes the embedding representation of nodes by type and code, and also initializes the embedding of edges based on version and type. Without edge embedding, NE-GCN degenerates into a traditional GCN. To explore the contribution of adding type embeddings to the patch representation, we conduct an ablation study using code-only embeddings to evaluate the impact on the three tasks. Table \ref{t6} shows the experimental results for different embedding approaches. We find that introducing type embedding significantly enhances the \name's performance in the three tasks, suggesting that type information plays a key role in the graph embedding and helps to strengthen the \name's ability to understand the code structure.

\subsubsection{\textbf{Effect of message passing module}}
To further explore the effectiveness of the NE-GCN module, we conduct comparative experiments between NE-GCN and the current mainstream GNN models, including GCN \cite{r28}, Graph Attention Network (GAT) \cite{r60}, and Gated Graph Neural Networks (GGNN) \cite{r61}. The results, as shown in Table \ref{t5}, show that NE-GCN achieves a significant performance improvement in all three tasks. As compared to the other best-performing GNN methods, NE-GCN achieves an increase of 4.83\%, 0.24\%, and 7.99\% in the F1-score, respectively. We attribute such improvements to the edge messaging layer introduced by NE-GCN, which captures richer information exchanges and dependencies and enhances the \name's ability to learn deeper features of the graph structure.

\begin{tcolorbox}[colframe=black,colback=gray!12,boxrule=1pt,left=3.5mm,right=3.5mm,top=1.5mm,bottom=1.5mm]
\textbf{Answer to RQ3: }The use of CPG, the design of merging CPGs, the embedded fusion of node types, and the design of NE-GCN in the \name model significantly enhance the performance of \name on the vulnerability-related tasks.
\end{tcolorbox}

\section{Threats to Validity}

Threats to validity arise from three main aspects: 

\textbf{Representativeness of data sources}: the vulnerability data in the constructed dataset all originates from fix cases that have been disclosed in open source vulnerability reports, which may not fully reflect those vulnerability fixes that have not been publicly disclosed. Although the data is derived from the widely recognized vulnerability database NVD, we will consider including more data sources in future studies to enhance the generalizability of our approach.

\textbf{Universality of programming languages}: we only conducted experiments on the dataset for Java, limiting the generalization of the research findings to implementations of OSSs in other programming languages (PLs). Having said that, the design of the module in \name is not limited to any specific PL, for instance, CPG serves as a common data structure representation for multiple languages. We plan to apply the \name model to datasets in other programming languages such as C/C++, and Python, in future research to further validate its performance.

\textbf{Implementation details of the baseline}: the specific implementation details of spi, the baseline method chosen for this study, are not publicly available. We conducted experiments based on the description in the original paper, strictly following its module design and parameter settings. For the parts not detailed in the original paper, we tried our best to adjust them to match the experimental results reported in the original paper. In addition, the node embedding part of GraphSPD is designed for C/C++ language, and we adapted it to Java language. However, due to differences in implementation details, uncertainties may still remain that lead to biases in the experimental results.

\section{Related Work}

\textbf{Deep-learning Model for Vulnerability Fix identification.} A large number of previous studies have proposed approaches to timely sensing silent vulnerability fixes \cite{r35}, \cite{r34}, \cite{r5}, \cite{r39}, \cite{r6}. Zhou et al. \cite{r39} designed a K-fold stacking classifier based on commit messages and bug reports to identify undisclosed vulnerabilities in open source reports. \cite{r34} trained two linear Support Vector Machine (SVM) classifiers to classify commit messages and code changes, respectively. A simple voting mechanism is used to combine the two classification results. SPI \cite{r6} and PatchRNN \cite{r5} utilize RNN to train a commit message classifier and a diff code classifier, respectively, and then composite the two results. HERMES \cite{r35} enriched the data sources by designing the commit-issue linking approach. In the real software maintenance process, information such as commit messages and issues are likely to be missing. Thus unlike the above studies, we identify possible vulnerability fixes by analyzing only the code changes themselves, while avoiding direct references to vulnerability information.

There exist a number of other vulnerability-fixes-aware approaches that focus on code changes \cite{r43}, \cite{r38}, \cite{r4}, \cite{r41}, \cite{r36}.  VCCFinder \cite{r38} is based on the SVM model to mark commits related to vulnerabilities. SPAIN \cite{r41} is a binary patch analysis framework that automatically identifies security patches. VulFixMiner \cite{r36} uses a pre-trained CodeBERT model to extract semantics from code changes to identify silent fixes. GraphSPD \cite{r4} designs a graph representation of patches and an end-to-end GNN model that determines whether a patch is security-relevant directly from the graph structure. MiDas \cite{r43} utilizes multiple levels of code change granularity and incorporates a code pre-training model to detect vulnerability fixes. Different from the above approaches, we further assess vulnerability fixes. We provide two critical information, the CWE type of the vulnerability and the severity of the vulnerability, which help software maintainers understand the vulnerability.

Colefunda \cite{r42} learns code representations from patches by employing function change data augmentation and comparative learning methods and provides explainable silent fix identification. While Colefunda has conceptual similarities with our approach, it focuses on function-level code representations and, since its tools are not publicly available, we are not able to perform a direct comparative analysis.

\textbf{Code representation of vulnerabilities.} Early studies \cite{r62}, \cite{r63}, \cite{r25}, \cite{r64} mostly regard the source code as flat serial data and rely on natural language processing techniques to represent the vulnerability code. This approach, while able to capture the semantic information of the code to some extent, ignores the inherent structural features of the code \cite{r47}, such as syntactic dependencies.
In order to better utilize the structural information of the code, several approaches \cite{r51}, \cite{r68}, \cite{r67}, \cite{r65}, \cite{r70}, \cite{r56}, \cite{r48}, \cite{r66}, \cite{r69}, \cite{r47} utilize the form of graphs to abstract the code for vulnerability related tasks.
Devign \cite{r47} utilizes GNN to iteratively propagate CPG information and learn rich semantic information in CPGs.
AMPLE \cite{r48} captures the long-distance dependencies of nodes through graph simplification and augmented graph representation learning.
The above study \cite{r47} used CPGs due to their rich semantic information. Our approach also employs CPGs and further mines its structural information to represent patches more efficiently.

\section{Conclusion}
In this work, we propose \name, a graph-based patch representation learning method for vulnerability fix identification and assessment. \name maps silent fix patches to a uniform graph structure MCPG, focusing on extracting and representing their structural features. NE-GCN is designed, which efficiently aggregates and passes the edge attributes and node attributes in the graph to comprehensively capture the structural features of vulnerability fixes.
To evaluate the effectiveness of \name, we construct a dataset containing 1068 vulnerability fixes and 1183 non-vulnerability fixes. Through experiments on three important tasks, including vulnerability fix identification, vulnerability types classification, and vulnerability severity classification, we comprehensively evaluate the performance of \name.
The evaluation results show that the \name model demonstrates excellent performance beyond the existing baseline model in all the above tasks. Compared with the best-performing baseline, the F1-score improves by 6.55\%, 12.57\%, and 14.98\% in each of the three tasks. The future works include expanding the vulnerability fix dataset, supplementing experiments with comparisons to traditional methods, and providing specific examples of the effectiveness of \name.

Our source code as well as the dataset are available at \href{https://github.com/han-mei/GRAPE}{https://github.com/han-mei/GRAPE}.

\section*{Acknowledgment}
This work was supported partially by the National Key R\&D Program of China under Grant No. 2018YFB1003901, the National Natural Science Foundation of China under Grant No.61572126 and No. 61872078, and the Cooperation Project with Huawei Technologies Co. Ltd under Grant No. YBN2016020009.

\bibliography{reference}
\end{document}